\documentclass[twocolumn,prl,reprint,amsmath,amssymb,showpacs]{revtex4}
\usepackage{amsmath,amssymb,graphicx,epstopdf}
\usepackage[usenames, dvipsnames]{color}

\begin{document}

\title{Observation of single-photon superradiance and the cooperative Lamb shift in an extended sample of cold atoms}

\author{S.J. Roof, K.J. Kemp and M.D. Havey}

\affiliation{Old Dominion University, Department of Physics, Norfolk, Virginia 23529 \
}

\author{I.M. Sokolov${}^{1,2}$}

\affiliation{$^{1}$Department of Theoretical Physics, State Polytechnic
University, 195251, St.-Petersburg, Russia \\ \small $^{2}$Institute for
Analytical Instrumentation, Russian Academy of Sciences, 198103, St.-Petersburg, Russia  }

\begin{abstract}
We report direct, time-resolved observations of single-photon superradiance in a highly extended, elliptical sample of cold $^{87}$Rb atoms. The observed rapid decay rate is accompanied by its counterpart, the cooperative Lamb shift.  The rate of the strongly directional decay, and the associated shift, scale linearly with the number of atoms, demonstrating the collective nature of the observed quantities.
\end{abstract}

\pacs{42.50.Gy,42.50.Nn,32.80.Qk,03.65.Yz}

\maketitle
Superradiance is the cooperative spontaneous emission of photons from a collection of atoms \cite{Gross1982}. Originally described by Dicke \cite{Dicke1954}, it is essentially a many-body effect arising from the synchronization of coupled radiating dipoles.  It is characterized by a substantially enhanced emission rate, compared to the decay rate of a single atom, and an associated cooperative frequency shift of the atomic resonance \cite{Friedberg1973,Manassah2012}.  There is extensive literature on superradiance in both atomic and condensed systems, see for examples \cite{Gross1982,MacGillivray1976,Gross1976,Skribanowitz1973,Rehler1971}. However, recent developments associated with single-photon \cite{Scully2006,Scully2007} super- and sub-radiance \cite{Guerin2016,Bienaime2012,Scully2015} have drawn new attention to this subject.  Cooperativity leads, for even dilute vapor samples, to optical depth dependent frequency shifts, line shape distortion, and suppression of signal size; these can have important impacts in quantum informatics \cite{Kimble1,Goban2015,DeOliveira2014}, quantum sensors \cite{Kitching} and optical lattice clocks \cite{Bromley,Nicholson1,Chang}.  Further, recent experimental observations and theoretical work have examined aspects of the fundamentally important cooperative Lamb shift \cite{Scully2009} in a variety of superradiant systems including X-ray emitters \cite{Rohlsberger2010}, a few body atomic ion system \cite{Meir2014}, and warm \cite{Keaveney2012} atomic gases.\\
\indent
An important component of recent studies is that the physical samples are considerably larger than the radiation wavelength $\lambda$, and thus fall outside the usual Dicke regime, where the radiating ensembles are much smaller than $\lambda$.   For such larger samples, the disordered spatial distribution of scatterers tends to randomize the relative phases of the emitters, discouraging coherent enhancement of the radiation.  One way to obtain strong coherent emission from the ensembles in this case is to prepare a timed-Dicke state \cite{Scully2009}.  This is accomplished by weak-field optical excitation resulting in uptake of a single photon by the ensemble.  The timed aspect reflects the phase distribution impressed on the atomic gas during optical excitation; it is this spatially varying phase that ensures coherent emission in directions close to that of the driving field.   A number of fascinating phenomena have been predicted, including complex many-body dynamics during the system decay and the role of virtual processes in them \cite{Scully5,Svidzinsky1}.\\
\indent
A unique feature of the cooperative shift and associated superradiant decay rate is that the $\emph{theoretically predicted}$ values of these quantities depend strongly on the average spatial distribution of scatterers \cite{Friedberg1973,Manassah2012}. For instance, the cooperative Lamb shift can be greatly enhanced in a particular lattice configuration \cite{Manassah2}, and yet vanish for an elliptical sample of the proper aspect ratio \cite{Manassah2012}. In the present case, we have used an elongated elliptical  sample for which the cooperative Lamb shift is predicted to be rather large, making it accessible to measurement in a relatively dilute atomic gas.\\
\indent
In this letter we present experimental studies, and associated analysis, of single photon superradiance in a dilute atomic gas. Our two most important results are (a) direct time domain measurements of superradiant emission from a spatially extended and low density cloud of cold $^{87}$Rb atoms, and (b) the corresponding cooperative Lamb shift of the atomic resonance studied. These quantities are each found to scale approximately linearly with the number of atoms, characteristic of a cooperative process in such cold atom systems \cite{Bienaime2}. The results are also found to be in good qualitative agreement with a vector light scattering simulation of the processes \cite{Sokolov2,Kupr}, and to correspond closely with predictions of a scalar coupled dipole model \cite{Skipetrov,Bellando}.\\
\indent
Cooperative interactions can be qualitatively understood as individual atoms interacting with the emitted fields of surrounding atoms, which results in observable quantities that differ from the single atom response. An effective model in investigating cooperative effects is the coupled dipole model \cite{Bienaime2011,Ruostekoski1997,Sutherland2016,Svidzinsky2010,Friedberg2008}, which gives the time evolution equations

\begin{equation}
\dot{ \beta }_j=-\frac{i}{2}\Omega_0e^{\textbf{k}_0\cdot \textbf{r}_j} + \left( i\Delta-\frac{\Gamma}{2} \right)\beta_j + \frac{i\Gamma}{2}\sum_{m\neq j}\beta_m \frac{e^{ik_ar_{jm}}}{k_ar_{jm}}
\label{CDM}
\end{equation}

\noindent Here, $\beta_j$ is the excited state amplitude of the $j$th atom, $\Omega_0$ the Rabi frequency of the incident probe beam, $\textbf{k}_0$ the probe wave number, $\Delta$ the probe detuning from resonance, $\Gamma$ the single atom decay rate, and $k_a$ the atomic resonance wave number. Atom interactions are represented by the summation in the last term which includes the long-range $1/r$ exponential kernel. This kernel serves to not only describe the real radiated field from one atom to the next but also the exchange of virtual photons which are responsible for the cooperative Lamb shift of the resonance line \cite{Friedberg2008,Scully2009}. One can extract expressions for the decay rate enhancement and frequency shift by considering that the sample is prepared in a timed-Dicke state $\left| +\right\rangle = \frac{1}{\sqrt{N}}\sum_{j=1}^{N} e^{i\textbf{k}_0 \cdot \textbf{r}_j} \left| j\right\rangle $ \cite{Scully2006}. Here the ket $\left| j\right\rangle$ represents a Fock state in which the atom labeled j is in the excited state, and all others are in the ground state.  That is, it is assumed that one atom is excited in the system but it is not known which one. Evaluation of the evolution of the timed Dicke state leads to the emergence of the quantities \cite{Bienaime2011,Friedberg2008}

\begin{subequations}
\begin{eqnarray}
\Gamma_N=\frac{\Gamma}{N}\text{Re} \left[ \sum_{j,m} \frac{e^{ik_ar_{jm}}}{ik_ar_{jm}} e^{-i\textbf{k}_0 \cdot \textbf{r}_{jm}} \right] \label{gamn}   \\
\Delta_N=\frac{\Gamma}{2N}\text{Im} \left[ \sum_{j, m \neq j} \frac{e^{ik_ar_{jm}}}{ik_ar_{jm}} e^{-i\textbf{k}_0 \cdot \textbf{r}_{jm}} \right] \label{deltan}
\end{eqnarray}
\end{subequations}

\noindent $\Gamma_N$ being the enhanced decay rate and $\Delta_N$ the cooperative shift. While the above expressions depend on the number of atoms they also have a strong dependence on the particular shape of the sample. This dependence can be quite substantial if the atoms distribution is highly extended along the direction of excitation \cite{Manassah2012}.

\begin{figure}
	\includegraphics[scale=.22]{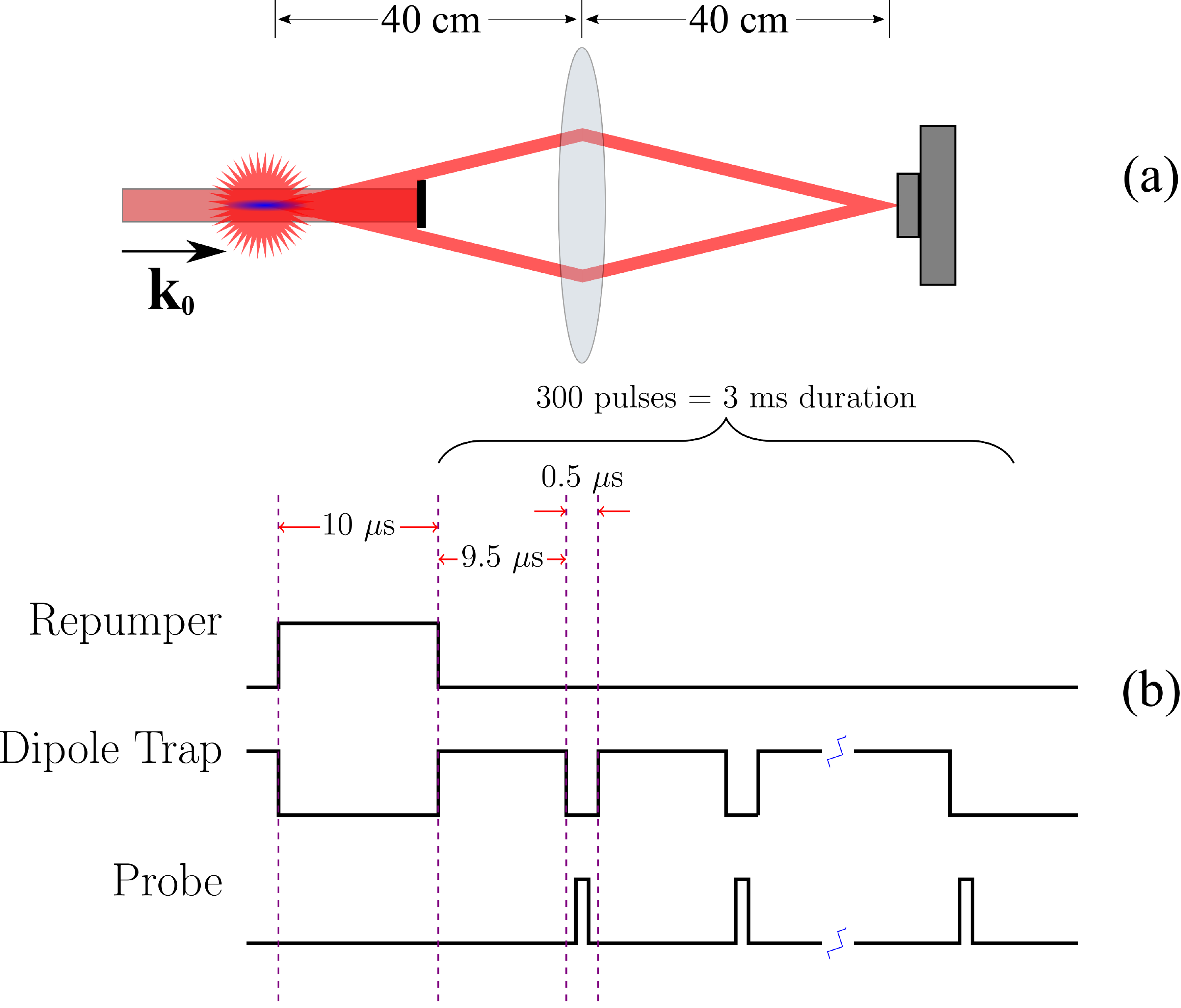}
	\caption{(a) The experimental configuration used. A near-resonant light beam is incident on a sample of $^{87}$Rb atoms prepared in an optical dipole trap. The nearly planar incident probe wave is blocked, while the forward emitted light is directed towards a detector by a 2f - 2f configuration (f$=20$ cm). (b) A schematic of the timing sequence used. After an initial thermalization stage, the dipole trapping beam is temporarily turned off and the atoms are pumped to the $F=2$ ground state with a variable length pulse from the repumping beams.}
	\label{fig1}
\end{figure}

We study cooperative effects of extended samples by preparing  $^{87}$Rb atoms in a multistep process by which atoms from a warm atomic vapor are cooled and loaded into a magneto-optical trap \cite{Metcalf}, and then transferred to a far-off-resonance dipole trap (FORT) \cite{Corwin2000,Balik2013}.  The characteristics of the sample result in an equilibrium temperature of $\sim 60\ \mu$K and maximum number of atoms near $10^5$ (peak density $6\times10^{12}$ atoms/cc).  The atomic sample is well described by a bi-Gaussian spatial distribution of atoms having Gaussian radii of $r_o$ = 2.7(1) $\mu$m and $z_o$ =  156(7) $\mu$m. A measurable quantity in the experiments is the number of atoms $N$ in the trap.  As the sample geometry, determined by the FORT parameters and atom temperature, is fixed in this experiment, varying $N$ is equivalent to changing either the optical depth or the peak density in the trap center.  The near-resonance laser beam used to probe the sample has a wavelength $\sim$ 780 nm, and is tuned in a range of $\pm$ 48 MHz about the 5$^2S_{1/2}$ $F = 2  \to$  5$^2P_{3/2}$ $F^{\prime} = 3$ hyperfine transition ($\Gamma=6.1$MHz). The probe beam has a beam waist $\sim 570$ $\mu$m and is linearly polarized; it is aligned along the long direction of the sample and is very nearly collimated such that it can be well-described as a plane wave.  The power is $\sim300$ nW, giving an on-resonance saturation parameter $s_o$ $\sim$ 0.03, such that we are well in the atomic linear response regime. Due to the highly directional emission of superradiance, the scattered light emerges from the sample in a lobe-shaped spatial profile having approximately a full angle $\lambda/r_0$ \cite{Gross1982}. This allows for ``mode-mismatching" between the incident probe beam and the emitted light such that upon exiting the sample chamber, the probe is occluded with a beam block and the forward emitted lobe largely passes around it. The forward scattered light is collected with a lens in a 2f - 2f configuration and focused onto a detector (see Fig.~\ref{fig1}(a)). It should be noted that due to the small Doppler width of the atoms ($\sim 100$ kHz) we expect there to be no contribution from anomalously fast decay which can occur in inhomogeneously broadened samples \cite{Chalony2011,Brewer1972}.\\
\indent
The scattered light is detected using two different schemes. The first method uses the technique of time-correlated single photon counting \cite{O'Connor,Simsarian1998} by which light signals are detected by a photomultiplier tube (PMT), sent through a constant-fraction discriminator to a time-to-amplitude converter, and counted with a multichannel analyzer. Due to the long duty cycle of building optical-dipole traps and the low intensity of the emitted light, the time-resolved measurements are made in a series of pulse trains \cite{Pellegrino2014a} to increase the efficiency of the experiment. Initially, atoms are loaded into the FORT into the $F=1$ ground state and after thermalization are variably pumped into $F=2$ ground state. The atoms are released from the dipole trap for 0.5 $\mu$s, probed with a short 15 ns pulse, and retrapped for another 9.5 $\mu$s (Fig.~\ref{fig1}(b)). This process is repeated for 300 iterations and has been checked to ensure that no significant heating or atom loss occurs. The probe pulse is derived from a lithium niobate intensity switch which is driven by the amplified output of a fast voltage comparator. The effective rise and fall time is $\sim$1 ns.\\
\begin{figure}
	\includegraphics[scale=.5]{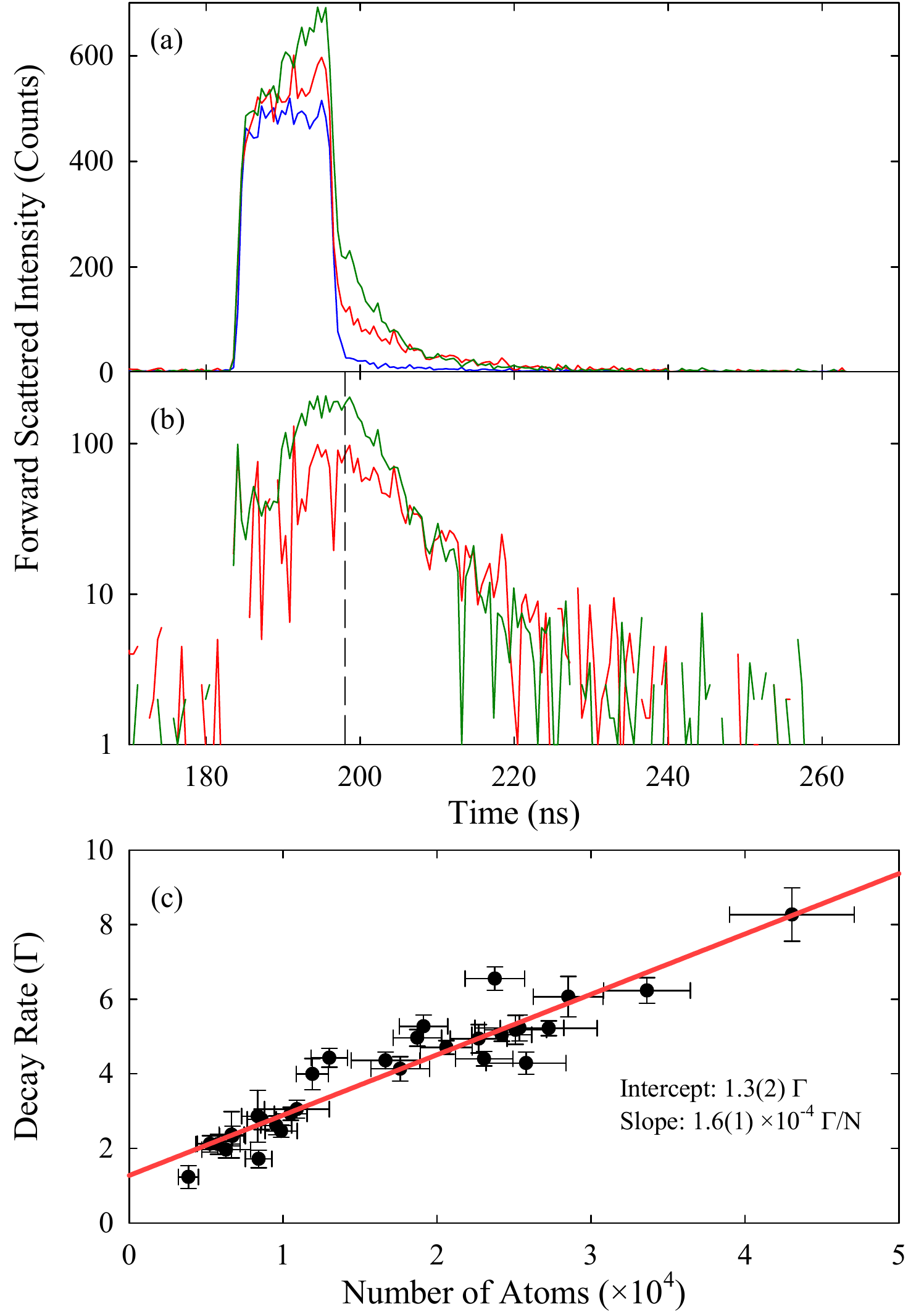}
	\caption{(a) Typical temporal responses for $\Delta$$=$$-2\Gamma$ for the bare probe pulse, N$\sim10000$ atoms, and N$\sim20000$ atoms (blue, red, and green, respectively). (b) Semi-log plot with the probe beam contribution subtracted from each signal. Exponential fits to the decay tails give 10.6(7) ns and 5.3(2) ns decays for N=10000 and N=20000 atoms, respectively. The black dotted line indicates where the probe beam shuts off. (c) The decay constants extracted from each fit are converted to a decay rate and plotted versus the number of atoms along with a linear fit to the data.}
	\label{fig2}
\end{figure}
\indent Representative measurements for the time-resolved decay are shown in Fig. \ref{fig2}(a). The signal size (in counts) corresponds to a total of $3.6 \times 10^{5}$ individual measurements consisting of 1200 realizations sampled 300 times each. After the probe beam is extinguished, the accumulated signal can be observed that decays more quickly for an increasing number of atoms. A small fraction of the incident probe light, present in the absence of the atom sample, does scatter into the detector; this fraction is attributed partially to diffraction from the beam block, but primarily is due to reflections from the entrance and exit windows which are not antireflection coated at the probe wavelength of 780 nm.   We subtract this parasitic level from the main signal when performing fits to the data. The resulting forward scattering signals are strongly linearly polarized in the same direction as the linearly polarized probe beam.  This observation reflects the absence of diffusely scattered light in the detected signals. Importantly, we find that the resulting decay rate increases approximately linear with the number of atoms and is insensitive to detuning.  We attribute the independence to two factors.  First, the superradiant states have an intrinsically large spectral width.  Second, the 15 ns probe laser beam pulses have a large transform limited bandwidth. The linear relationship between the decay rate and the number of atoms is consistent with the theory of single photon superradiance, for which evaluation of Eq. (\ref{gamn}) in the continuous limit gives the relation $\Gamma_N=\Gamma+\alpha (N-1)\Gamma$ \cite{Scully2007,Bienaime2011,Sutherland2016}. Here $\alpha$ is a factor that depends on the particular geometric shape of the sample \cite{Courteille2010},

\begin{eqnarray}
\alpha&=&\frac{\sqrt{\pi}}{4\sigma\sqrt{{\eta}^2-1}} \exp\left[\frac{\sigma^2}{\eta^2-1}\right] \times \nonumber\\ &&\left\{\text{erf}\left[\frac{\sigma}{\sqrt{\eta^2-1}}\left(2\eta^2-1\right)\right] - \text{erf}\left[\frac{\sigma}{\sqrt{\eta^2-1}}\right]\right\},\nonumber\\
\label{alpha}
\end{eqnarray}

\noindent where $\sigma=k_0r_0$ and $\eta=\frac{z_0}{r_0}$. In Fig. \ref{fig2}(c) decay rates extracted from all runs are plotted versus the number of atoms and fit to a straight line.  This results in a slope of $1.6(1) \times 10^{-4}$ $\Gamma $/$N$ and an intercept very near the single atom decay rate. Evaluating Eq. (\ref{alpha}) results in a slope of 2.4(3)$\times 10^{-4}$ $\Gamma $/$N$, the uncertainty coming from experimental sample size uncertainties. This difference is not surprising, as that analysis is based on a two-level system where level degeneracy and photon polarization are not taken into account.\\
\begin{figure}
	\includegraphics[scale=.5]{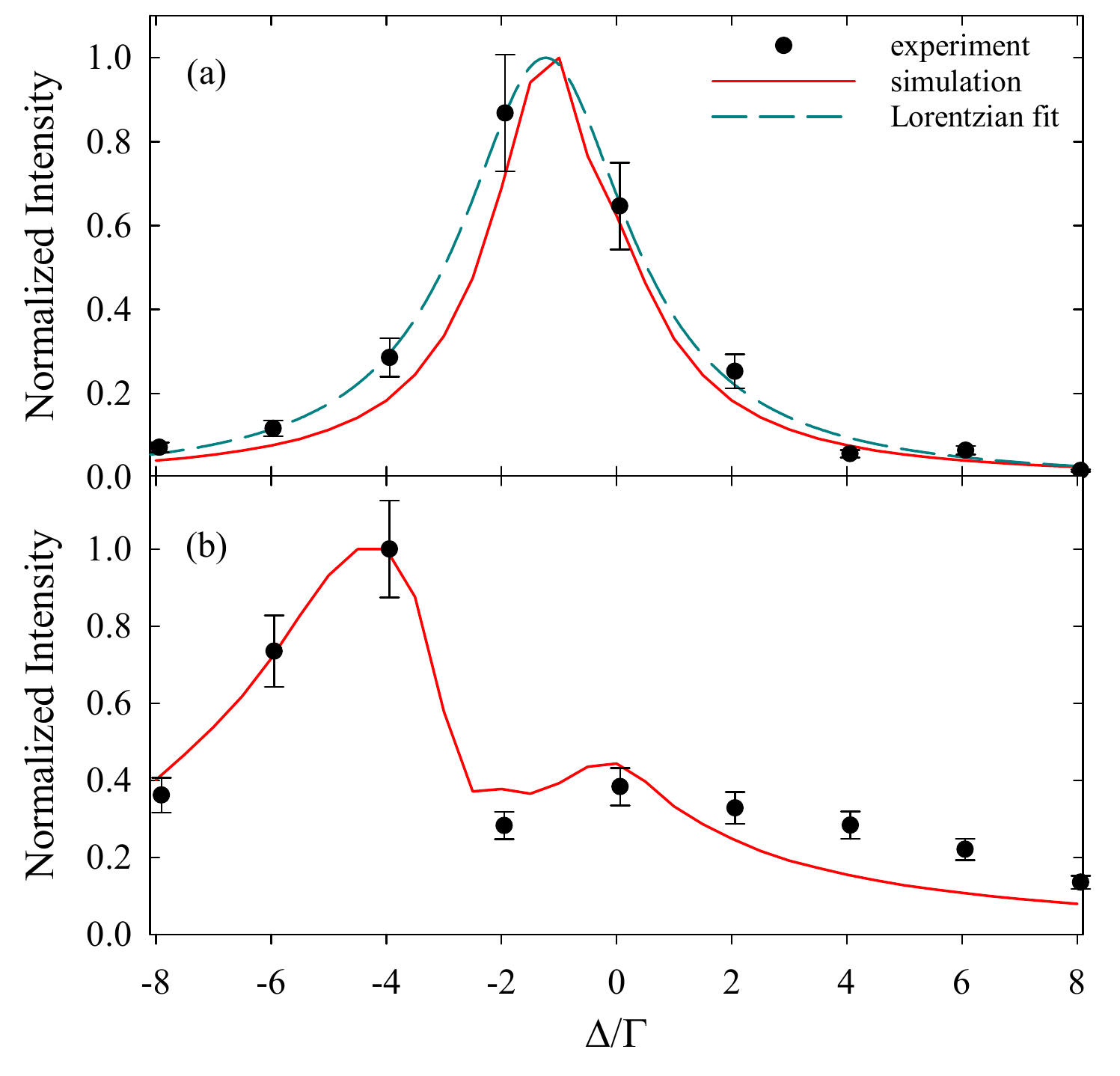}
	\caption{(a) Experimental data (black dots) for an atom number of $\sim14$k atoms. A Lorentzian curve is fit to the data (dashed blue curve) and overlayed with a simulation of 2500 atoms with sample geometry $r_0=1.19$ $\mu$m and $z_0=71.4$ $\mu$m (red curve). To make a uniform comparison, each data set has been normalized.  (b) An atom number of $\sim53$k atoms (black dots) overlayed with a simulation of 10000 atoms with the same sample geometry as (a).}
	\label{fig3}
\end{figure}
\indent
In the second stage of the experiment the sample is probed only once per trap realization with a temporally longer $10$ $\mu$s pulse such that the pulse transform limited width is much smaller, and the spectral response of the atomic system can be resolved. As time resolution of the forward scattered light is not essential to this part of the experiment, the PMT is replaced with a charge-coupled device (CCD) camera and the emitted light is collected over the full duration of the probing cycle.  It is checked that no significant atom loss occurs during probing due to optical pumping or any other mechanism. Even with such a low-rate duty cycle, the high quantum efficiency of the CCD ($> 90\%$) and much longer excitation pulse allows for sufficient signal collection. Representative data for the steady state regime are given in Fig. \ref{fig3}. Here it can be seen that the response spectral width is broadened, consistent with decay rates extracted from the time resolved data. It is also observed that the center of the response peak has been shifted to a lower frequency. For lower atom numbers the spectral response is symmetric within the uncertainties of the data.  These response curves have been used to determine the centroid of the measured response for a range of atom numbers. The resulting dependence of the shift on atom number is shown in Fig. \ref{fig4} along with a linear fit through the data.\\
\indent
To compare calculations with experiment, we numerically solve Eq. (\ref{CDM}) for many realizations of the atomic distribution  (subject to the average bi-Gaussian sample shape) which is reflected in the atom locations $\textbf{r}_{j}$.  This permits calculation of the average value of the excited state amplitudes $\beta_j$, from which observables may be readily obtained.
The sample in simulation has to be rescaled due to computational memory limitations. This is accomplished by choosing $N=10$k atoms to correspond to the experimental sample of $N=53$k atoms, keeping the aspect ratio $\eta$ the same as in experiment, and rescaling the system dimension to produce the same optical depth estimated in the experiment. Here, the optical depth is used as an approximate scaling factor to compare experiment and theory. The lower atom number comparison in Fig. \ref{fig3}(a) is then achieved by reducing the number of atoms by a factor of 4. The validity of the scaling approach was established for those cases where the number of atoms is small enough to do direct comparisons.   Numerical calculation of Eq. (\ref{deltan}), although for a two-level atom, produces a shift on the same order suggesting that the origin is the cooperative Lamb shift. Friedberg et al. \cite{Friedberg2010} have considered such a sample shape for an continuous distribution of two-level atoms but in the limiting case that $\eta >> \sigma$. There, they also predict a shift on the same order for an ellipsoid but their result slightly underestimates what is found in Eq. (\ref{deltan}), as the ratio $\eta/\sigma\sim3$ in experiment and calculation. Observing a shift as large as the one measured may seem surprising given the low density of the sample. However, it supports the predicted importance of sample geometry in determining the mode coupling between the atom and photon field \cite{Manassah2012}.\\
\indent
Also shown in Fig. \ref{fig3} is the measured and simulated response (using Eq. (\ref{CDM})) for a much larger atom number.   There we see that the shift to lower frequencies of the main response continues the pattern observed at lower atom number.  However, there is substantial distortion of the spectral profile in this case, indicating that at larger atom number we are entering a different physical regime.  Performing simulations using Eq. (\ref{CDM}) even at higher atom number gives good agreement between experiment and theory.  The observed spectral distortion is plausibly due to complex pulse propagation effects in the optically deep, anisotropic, and inhomogeneous atomic sample.\\
\begin{figure}
	\includegraphics[scale=.5]{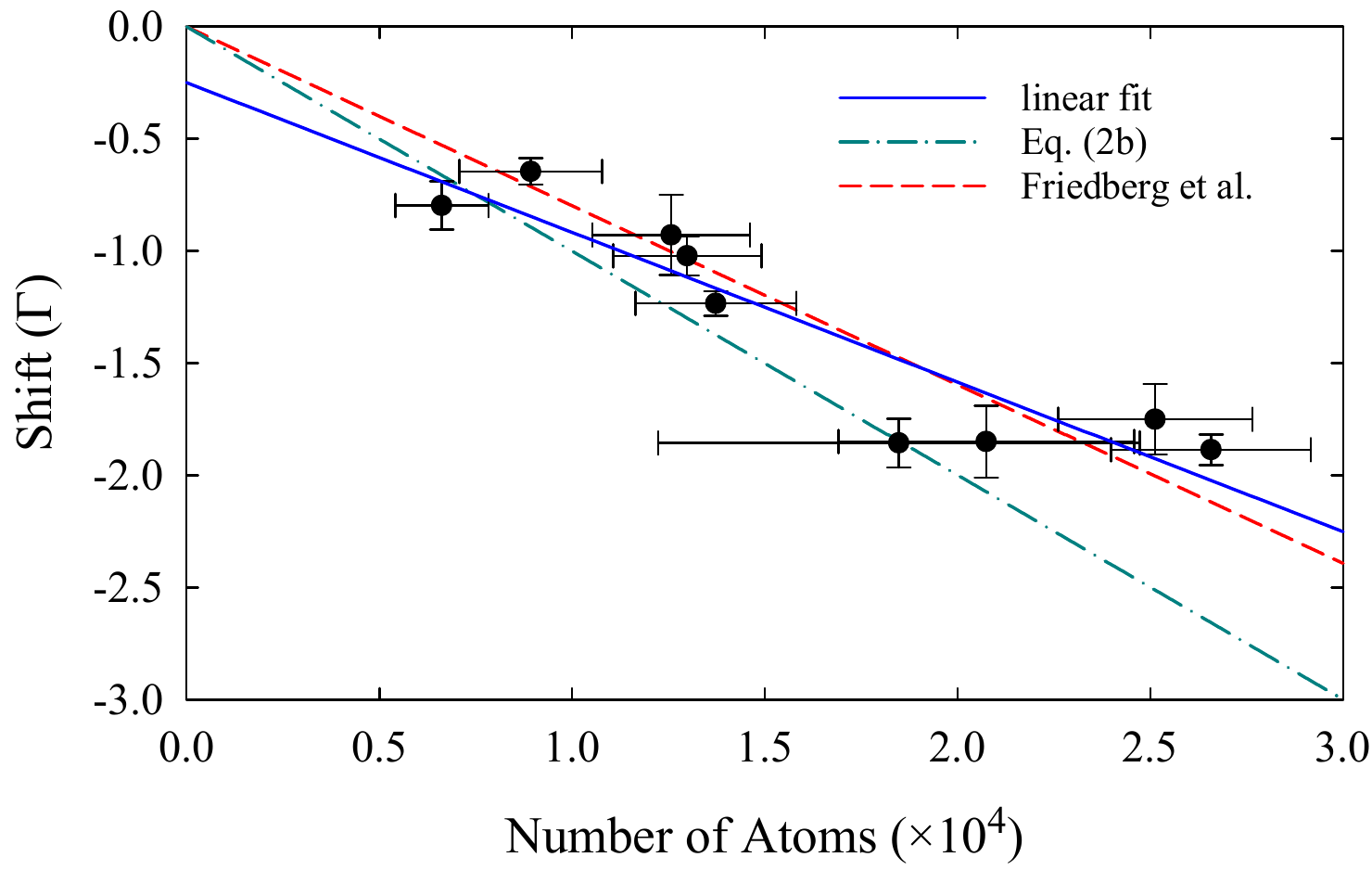}
	\caption{Plot of the shift versus the number of atoms. The shift is extracted by fitting the data to a Lorentzian and is limited to the spectra that show no significant spectral distortion. A line is fit to the observed shift and compared to the theoretical values as predicted by Eq. (\ref{deltan}) and \cite{Friedberg2010}.}
	\label{fig4}
\end{figure}
\indent
In conclusion, we have directly measured the superradiant emission and associated cooperative frequency shift for a highly extended sample of cold atoms. The superradiant emission is observed in the time domain as a rapid emission of light from the sample and as line broadening in the steady-state regime. The measured cooperative frequency shift is quite large and is due to the particular theoretical dependence that the shift has on sample shape. The superradiant rate and the cooperative frequency shifts agree well with theory and simulation. Finally, for a larger number of atoms the spectral response becomes somewhat distorted, suggesting that complex propagation effects come into play.  Each of these factors may have important impacts in cold-atom realizations of single photon storage, quantum sensors, or next generation atomic clocks. The theoretically predicted role of sample geometry \cite{Manassah2012,Manassah2} is consistent with our measurements of the frequency shift and superradiant decay rate; future experimental study of the shape dependence may usefully minimize the frequency shifts in practical applications.

We appreciate financial support by the National Science Foundation (Grant Nos. NSF-PHY-0654226 and NSF-PHY-1068159) and by the Russian Foundation for Basic Research (Grant No. RFBR-15-02-01013. We also thank R. T. Sutherland for fruitful discussions.


\begin{thebibliography}{99}

\bibitem{Gross1982} M. Gross and S. Haroche, Phys. Rep. \textbf{93}, 301 (1982).

\bibitem{Dicke1954} R. H. Dicke, Phys. Rev. \textbf{93}, 99 (1954).

\bibitem{Friedberg1973} R. Friedberg, S. R. Hartmann, J. T. Manassah, Phys. Rep. \textbf{7}, 101 (1973).

\bibitem{Manassah2012} J. T. Manassah, Adv. Opt. Photonics \textbf{4}, 108 (2012).

\bibitem{MacGillivray1976} J. C. MacGillivray, and M. S. Feld, Phys. Rev. A \textbf{14}, 1169 (1976).

\bibitem{Gross1976} M. Gross, C. Fabre, P. Pillet, and S. Haroche, Phys. Rev. Lett. \textbf{36}, 1035 (1976).

\bibitem{Skribanowitz1973} N. Skribanowitz, I. P. Herman, J. C. MacGillivray, and M. S. Feld, Phys. Rev. Lett. \textbf{30}, 309 (1973).

\bibitem{Rehler1971} N. Rehler, and J. Eberly, Phys. Rev. A \textbf{3}, 1735 (1971).

\bibitem{Scully2006} M. O. Scully, E. S. Fry, C. H. Raymond Ooi, and K. W\`odkiewicz, Phys. Rev. Lett. \textbf{96}, 010501 (2006).

\bibitem{Scully2007} Marlan O. Scully, Laser Physics \textbf{17}, 635 (2007).

\bibitem{Guerin2016} W. Guerin, M. O. Ara{\'{u}}jo, R. Kaiser, Phys. Rev. Lett. \textbf{116}, 083601 (2016).

\bibitem{Bienaime2012} T. Bienaim{\'{e}}, N. Piovella, R. Kaiser, Phys. Rev. Lett. \textbf{108}, 123602 (2012).

\bibitem{Sokolov2} I.M. Sokolov, D.V. Kupriyanov, and M.D. Havey, JETP \textbf{112}, 246 (2011).

\bibitem{Kupr} I.M. Sokolov, M.D. Kupriyanova, D.V. Kupriyanov, and M.D. Havey, Phys. Rev. A \textbf{79}, 053405 (2009).

\bibitem{Scully2015} M. O. Scully, Phys. Rev. Lett. \textbf{115}, 243602 (2015).

\bibitem{Kimble1} H.J.Kimble, Nature \textbf{453}, 1023 (2008).

\bibitem{Goban2015} A. Goban, C. L. Hung, J. D. Hood, S. P. Yu, J. A. Muniz, O. Painter, and H. J. Kimble, Phys. Rev. Lett. \textbf{115}, 063601 (2015).

\bibitem{DeOliveira2014} R. A. de Oliveira, M. S. Mendes, W. S. Martins, P. L. Saldanha, J. W. R. Tabosa, and D. Felinto, Phys Rev. A \textbf{90}, 023848 (2014).

\bibitem{Kitching} John Kitching, Svenja Knappe, and Elizabeth A. Donley, IEEE Sensors Journal, Vol. 11, 1749 (2011).

\bibitem{Bromley} S. L. Bromley, B. Zhu, M. Bishof, X. Zhang, T. Bothwell, J. Schachenmayer, T. L. Nicholson, R. Kaiser, S. F. Yelin, M. D. Lukin, A. M. Rey, J. Ye, Nature Commun. \textbf{7}, 11039 (2016).

\bibitem{Nicholson1}  T.L. Nicholson, S.L. Campbell, R.B. Hutson, G.E. Marti, B.J. Bloom, R.L. McNally,	W. Zhang, M.D. Barrett,	 M.S. Safronova,	G.F. Strouse, W.L. Tew, and J. Ye, Nat. Commun. \textbf{6}, 6896 (2015).
    
\bibitem{Chang} D.E. Chang, Jun Ye, and M.D. Lukin, Phys. Rev. A \textbf{69}, 023810 (2004).

\bibitem{Scully2009} M. O. Scully, Phys. Rev. Lett. \textbf{102}, 143601 (2009).

\bibitem{Rohlsberger2010} R. R{\"{o}}hlsberger, K. Schlage, B. Sahoo, S. Couet, and R. R{\"{u}}ffer, Science \textbf{238}, 1187770 (2010). 
\bibitem{Meir2014} Z. Meir, O. Schwartz, E. Shahmoon, D. Oron, and R. Ozeri, Phys. Rev. Lett. \textbf{113}, 193002 (2014).

\bibitem{Keaveney2012} J. Keaveney, A. Sargsyan, U. Krohn, I. G. Hughes, D. Sarkisyan, and C. S. Adams, Phys. Rev. Lett. \textbf{108}, 173601 (2012).

\bibitem{Scully5} Marlan O. Scully and Anatoly A. Svidzinsky, Phys. Lett. A \textbf{373}, 1283 (2009).

\bibitem{Svidzinsky1} Anatoly A. Svidzinsky and Marlan O. Scully, Optics Comm. \textbf{282}, 2894 (2009).

\bibitem{Manassah2} J.T. Manassah, Phys. Lett. A \textbf{374}, 1985 (2010).

\bibitem{Bienaime2} Tom Bienaim$\acute{e}$, Romain Bachelard, Nicola Piovella, and Robin Kaiser, Fortschr. Phys. \textbf{61}, 377 (2013).

\bibitem{Skipetrov} S.E. Skipetrov and I.M. Sokolov, Phys. Rev. Lett. \textbf{112}, 023905 (2014).

\bibitem{Bellando} L. Bellando, A. Gero, E. Akkermans, and R. Kaiser, Phys. Rev. A \textbf{90}, 063822 (2014).

\bibitem{Bienaime2011} T. Bienaim{\'{e}}, M. Petruzzo, D. Bigerni, N. Piovella, R. Kaiser, J. Mod. Opts. \textbf{21}, 594911 (2011).

\bibitem{Sutherland2016} R. T. Sutherland and F. Robicheaux, Phys. Rev. A \textbf{93}, 023407 (2016).

\bibitem{Ruostekoski1997} J. Ruostekoski and Juha Javanainen, Phys. Rev. A \textbf{55}, 513 (1997).

\bibitem{Svidzinsky2010} A. A. Svidzinsky, J. T. Chang, and M. O. Scully, Phys. Rev. A \textbf{81}, 053821 (2010).

\bibitem{Friedberg2008} R. Friedberg and J. T. Manassah, Phys. Lett. A \textbf{372}, 2514 (2008).

\bibitem{Metcalf} H. J. Metcalf, P. van der Straten, \textit{Laser Cooling and Trapping}, (Springer-Verlag, New York, 1999).

\bibitem{Corwin2000} S. J. M. Kuppens, K. L. Corwin, K. W. Miller, T. E. Chupp, and C. E. Wieman, Phys. Rev. A \textbf{62}, 013406 (2000).

\bibitem{Balik2013} S. Balik, A.L. Win, M.D. Havey, I.M. Sokolov and D.V. Kupriyanov, Phys. Rev. A \textbf{87}, 053817 (2013).

\bibitem{Chalony2011} M. Chalony, R. Pierrat, D. Delande, and D. Wilkowski, Phys. Rev. A \textbf{84}, 011401 (2011).

\bibitem{Brewer1972} R. G. Brewer and R. L. Shoemaker, Phys. Rev. A \textbf{6}, 2001 (1972).

\bibitem{O'Connor} D. V. O'Connor and D. Phillips, \emph{Time-correlated Single Photon Counting}, (Academic Press, London, UK, 1984).

\bibitem{Simsarian1998} J. E. Simsarian, L. A. Orozco, G. D. Sprouse, and W. Z. Zhao, Phys. Rev. A \textbf{57}, 2448 (1998).

\bibitem{Pellegrino2014a} J. Pellegrino, R. Bourgain, S. Jennewein, Y.R.P. Sortais, A. Browaeys, S.D. Jenkins and J. Ruostekoski, Phys. Rev. Lett. \textbf{113}, 133602 (2014).

\bibitem{Courteille2010} Ph. W. Courteille, S. Bux, E. Lucioni, K. Lauber, T. Bienaim{\'{e}}, R. Kaiser, N. Piovella, Eur. J. Phys. D \textbf{58}, 69 (2010).

\bibitem{Friedberg2010} R. Friedberg and J. T. Manassah, Phys. Rev. A \textbf{81}, 063822 (2010).


\end{thebibliography}
\end{document}